\documentclass{INTERSPEECH2024}

\interspeechcameraready

\usepackage{multicol}
\usepackage{multirow}
\usepackage{color, colortbl}
\usepackage{xcolor}
\colorlet{eng}{blue!10}
\colorlet{cmn}{teal!10}
\colorlet{multi}{yellow!10}
\colorlet{euro}{orange!10}
\usepackage{pifont}
\usepackage{hyperref}
\usepackage{graphicx}
\usepackage{subcaption}
\usepackage{tablefootnote}
\usepackage{bbding}
\title{TokSing: Singing Voice Synthesis based on Discrete Tokens}

\makeatletter
\def\thanks#1{\protected@xdef\@thanks{\@thanks
        \protect\footnotetext{#1}}}
\makeatother

\name[affiliation={1}]{Yuning}{Wu}
\name[affiliation={2}]{Chunlei}{Zhang}
\name[affiliation={3}]{Jiatong}{Shi}
\name[affiliation={1}]{Yuxun}{Tang}
\name[affiliation={2}]{Shan}{Yang}
\name[affiliation={1}]{Qin}{Jin*}\thanks{ *Corresponding Author.}

\address{$^{1}$ Renmin University of China,
        $^{2}$ Tencent AI Lab,
        $^{3}$ Carnegie Mellon University,
}

\email{\{yuningwu, qjin\}@ruc.edu.cn, cleizhang@global.tencent.com, jiatongs@cs.cmu.edu}

\usepackage[
backend=biber,
style=ieee,
citestyle=numeric-comp,
maxbibnames=10,
maxcitenames=20,
doi=false,isbn=false,url=false,eprint=false
]{biblatex}

\defbibheading{bibliography}[\refname]{}

\DeclareSourcemap{
	\maps[datatype=bibtex, overwrite=true]{
		\map{
			\step[fieldsource=booktitle,
			match=\regexp{.*Interspeech.*},
			replace={Proc. Interspeech}]
			\step[fieldsource=journal,
			match=\regexp{.*INTERSPEECH.*},
			replace={Proc. Interspeech}]
			\step[fieldsource=booktitle,
			match=\regexp{.*ICASSP.*},
			replace={Proc. ICASSP}]
			\step[fieldsource=booktitle,
			match=\regexp{.*icassp_inpress.*},
			replace={Proc. ICASSP (in press)}]
			\step[fieldsource=booktitle,
			match=\regexp{.*Acoustics,.*Speech.*and.*Signal.*Processing.*},
			replace={Proc. ICASSP}]
			\step[fieldsource=booktitle,
			match=\regexp{.*International.*Conference.*on.*Learning.*Representations.*},
			replace={Proc. ICLR}]
			\step[fieldsource=booktitle,
			match=\regexp{.*International.*Conference.*on.*Computational.*Linguistics.*},
			replace={Proc. COLING}]
			\step[fieldsource=booktitle,
			match=\regexp{.*SIGdial.*Meeting.*on.*Discourse.*and.*Dialogue.*},
			replace={Proc. SIGDIAL}]
			\step[fieldsource=booktitle,
			match=\regexp{.*International.*Conference.*on.*Machine.*Learning.*},
			replace={Proc. ICML}]
			\step[fieldsource=booktitle,
			match=\regexp{.*North.*American.*Chapter.*of.*the.*Association.*for.*Computational.*Linguistics:.*Human.*Language.*Technologies.*},
			replace={Proc. NAACL}]
			\step[fieldsource=booktitle,
			match=\regexp{.*Empirical.*Methods.*in.*Natural.*Language.*Processing.*},
			replace={Proc. EMNLP}]
			\step[fieldsource=booktitle,
			match=\regexp{.*Association.*for.*Computational.*Linguistics.*},
			replace={Proc. ACL}]
			\step[fieldsource=booktitle,
			match=\regexp{.*Automatic.*Speech.*Recognition.*and.*Understanding.*},
			replace={Proc. ASRU}]
			\step[fieldsource=booktitle,
			match=\regexp{.*Spoken.*Language.*Technology.*},
			replace={Proc. SLT}]
			\step[fieldsource=booktitle,
			match=\regexp{.*Speech.*Synthesis.*Workshop.*},
			replace={Proc. SSW}]
			\step[fieldsource=booktitle,
			match=\regexp{.*workshop.*on.*speech.*synthesis.*},
			replace={Proc. SSW}]
			\step[fieldsource=booktitle,
			match=\regexp{.*Advances.*in.*neural.*information.*processing.*},
			replace={Proc. NeurIPS}]
			\step[fieldsource=booktitle,
			match=\regexp{.*Advances.*in.*Neural.*Information.*Processing.*},
			replace={Proc. NeurIPS}]
			\step[fieldsource=booktitle,
			match=\regexp{.*Workshop.*on.* Applications.* of.* Signal.*Processing.*to.*Audio.*and.*Acoustics.*},
			replace={Proc. WASPAA}]
			\step[fieldsource=publisher,
			match=\regexp{.+},
			replace={{}}]
			\step[fieldsource=month,
			match=\regexp{.+},
			replace={{}}]
			\step[fieldsource=location,
			match=\regexp{.+},
			replace={{}}]
			\step[fieldsource=address,
			match=\regexp{.+},
			replace={{}}]
			\step[fieldsource=organization,
			match=\regexp{.+},
			replace={{}}]
		}
	}
}


\begin{document}

\maketitle
 
\begin{abstract}
Recent advancements in speech synthesis witness significant benefits by leveraging discrete tokens extracted from self-supervised learning (SSL) models. Discrete tokens offer higher storage efficiency and greater operability in intermediate representations compared to traditional continuous Mel spectrograms. However, when it comes to singing voice synthesis~(SVS), achieving higher levels of melody expression poses a great challenge for utilizing discrete tokens. In this paper, we introduce \textbf{\textit{TokSing}}, a discrete-based SVS system equipped with a token formulator that offers flexible token blendings. 
We observe a melody degradation during discretization, prompting us to integrate a melody signal with the discrete token and incorporate a specially-designed melody enhancement strategy in the musical encoder. Extensive experiments demonstrate that our TokSing achieves better performance against the Mel spectrogram baselines while offering advantages in intermediate representation space cost and convergence speed.

\end{abstract}

\section{Introduction}
\label{sec:intro}

Singing Voice Synthesis (SVS) aims to generate vocal sounds given music scores with melody and lyrics. 
Traditional SVS systems \cite{Lu2020XiaoiceSingAH, wang2022xiaoicesing, Liu2021DiffSingerSV, Shi2020SequenceToSequenceSV, Wang2022SingingTacotronGD, gu2021bytesing} primarily focus on enhancing acoustic models to generate Mel spectrograms from scores, which are then converted into waveforms by vocoders \cite{kong2020hifi, Lee2022BigVGANAU}. 
Recently, there has been a growing trend towards using discrete tokens, a representation with superior storage efficiency and controllability, for speech understanding and generation tasks \cite{Chang2023ExplorationOE, Yang2023TowardsUS, Zhang2023SpeechGPTEL, Kharitonov2023SpeakRA}. Discrete tokens can be obtained from raw audio through vector quantization \cite{Zeghidour2021SoundStreamAE, Defossez2022HighFN, Kumar2023HighFidelityAC}
 or generated by clustering \cite{MacQueen1967SomeMF} on hidden embeddings of SSL models \cite{Hsu2021HuBERTSS, baevski2020wav2vec, Chen2021WavLMLS, Baevski2019vqwav2vecSL} pretrained on large-scale audios. 

Different from speech processing, singing adds the complexity of melodic expression on top of speech, requiring vocal sound that meets the musical score requirements and delivers high-quality listening experiences. 
Therefore, applying discrete tokens in SVS faces some unique challenges. 
Firstly, limited by copyright restrictions and strict recording environments, there is currently no dedicated singing SSL model, and existing SSL models contain scarce singing data~\cite{Shi2020SequenceToSequenceSV, guo2022singaug, Shi2024SingingVD}. Therefore, setting suitable tokens for singing synthesis remains challenging.
Secondly, although lyrics are inherently discrete among the information encompassed within singing, the melody requires more refined expression, for example, the fundamental frequency of the same note can vary delicately between the frames it covers, especially in cases involving sustained notes, high pitches, vibratos, and other techniques requiring advanced vocal skills~\cite{yi2019singing}. Therefore, there is a risk of losing the acoustic details of the melody during the process of discretization.
Consequently, constructing a discrete token-based SVS system that meets the demands of melody expression poses a challenge. 

In this study, we focus on addressing the above two challenges: finding suitable token formulations for singing and constructing a discrete token-based SVS system that meets the demands of melody expression. 
Firstly, we introduce a method for formulating tokens tailored to singing tasks and provide various flexible formulation strategies.
Due to the diversity in pre-training tasks and datasets, different SSL models can offer valuable insights into singing semantics and acoustics, potentially providing complementary benefits to each other.
Additionally, drawing from past research \cite{Yang2021SUPERBSP, Chen2021WavLMLS, shi2023multi}, it is evident that the influence of various intermediate layers in an SSL model differs across downstream tasks. This suggests a distribution of diverse knowledge across the layers. 
Based on the above reasons, we propose a token formulator that allows token blending across different models and layers. 
Secondly, for the discrete SVS system, we incorporate melody control signals to enhance the generated melody expression. 
Finally, combining the above methods, we propose a new SVS framework, namely \textbf{\textit{TokSing}}, using discrete token sequences and melody-oriented signals as system intermediates. 
Our main contributions include: (1) We introduce a token formulator for training and provide multiple token formulations, offering flexibility in token sourcing and blending. (2) We propose a discrete token based SVS framework, TokSing, which achieves melody expression enhancement by integrating melody control signals to offset the loss of melodic intricacies in tokens. (3) Extensive experiments in both single-singer and multi-singer scenarios demonstrate that our proposed TokSing framework achieves better performance with lower storage cost and higher convergence speed.

\section{Method}
\label{sec:method}

\begin{figure}[t]
	\centering
	\includegraphics[width=0.9\columnwidth]{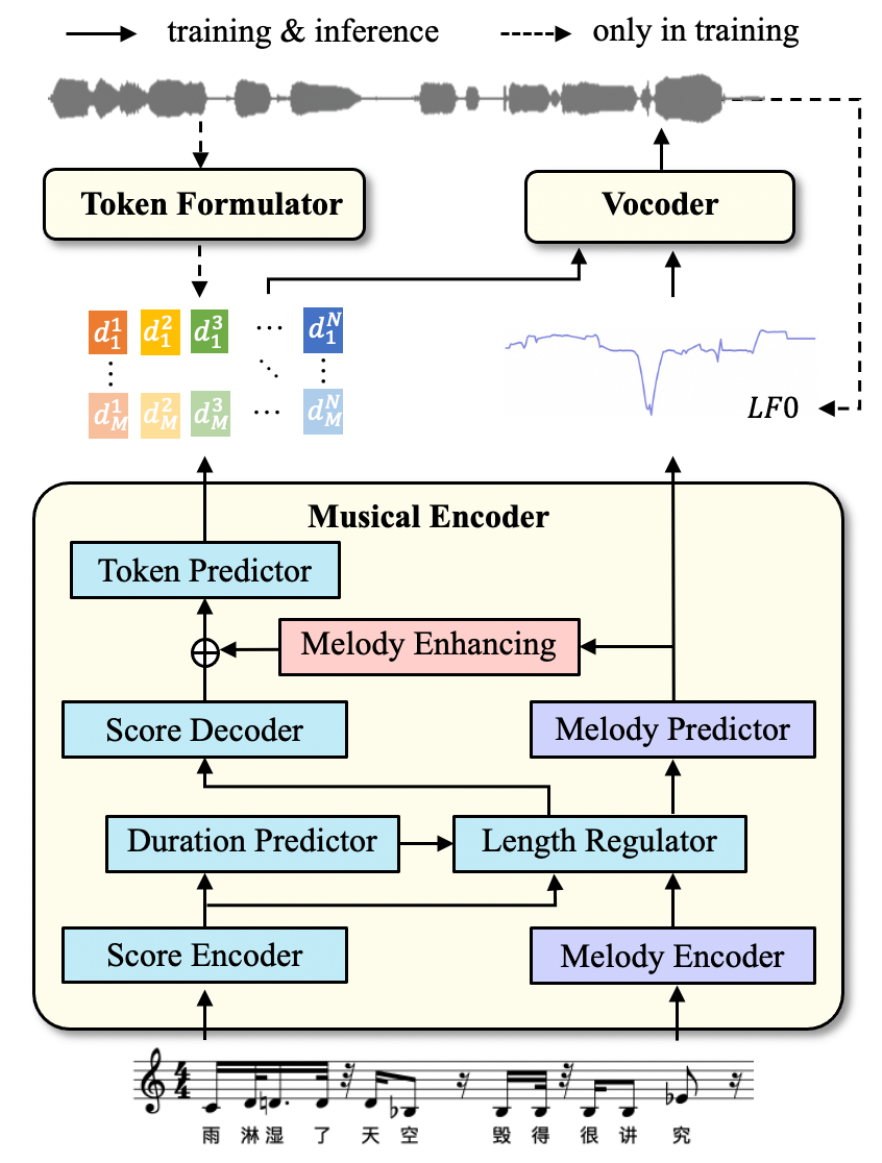}
        \vspace{-8pt}
	\caption{\small Discrete-based SVS system architecture. The system contains three parts: a token formulator, a musical encoder and a vocoder. $LF0$, the logarithm of the fundamental frequency melody signal, serves as the melody signal. The ablations of melody prediction modules (purple blocks) and melody enhancing module (pink block) are discussed in Section~\ref{ssec: melody}. $d_i^j$ represents discrete token.}
	\label{fig:system}
\end{figure}
\vspace{-4pt}

Figure~\ref{fig:system} illustrates the overall framework of our discrete-based SVS system, containing: 1) A token formulator extracts the hidden embeddings of the SSL model and quantizes them into tokens through clustering, with enhancements applied for better predictions. 2) A musical encoder inputs the music score and predicts target tokens and melody signals, where enhancement is applied for better predictions; 3) A vocoder converts tokens and melody signals into singing waveforms. Details of the three components are presented in the following subsections.

\subsection{Token Formulation}
\label{ssec: token formulation}
There are typically two ways to formulate discrete tokens. One way is to use a Variational Quantized Variational Autoencoder~(VQ-VAE) \cite{Oord2017NeuralDR} to derive discrete representations from raw audio. \cite{Zeghidour2021SoundStreamAE} introduces a residual vector quantizer (RVQ) and can reconstruct audios from multi-layer Codec tokens with Codec decoder.
The other way involves quantizing tokens through clustering the hidden embeddings of SSL models. The pre-training task and the corpus of these models directly influence token generation. Previous research on speech understanding tasks \cite{Chen2021WavLMLS, shi2024multiresolution, shi2024Ommm} suggests that different hidden layers contain diverse information suitable for various downstream tasks, with shallow layers focusing more on identity and deeper layers on content-related features. 
We leverage and compare these tokens for SVS task. 

\begin{figure}[t]
	\centering
	\includegraphics[width=0.7\columnwidth]{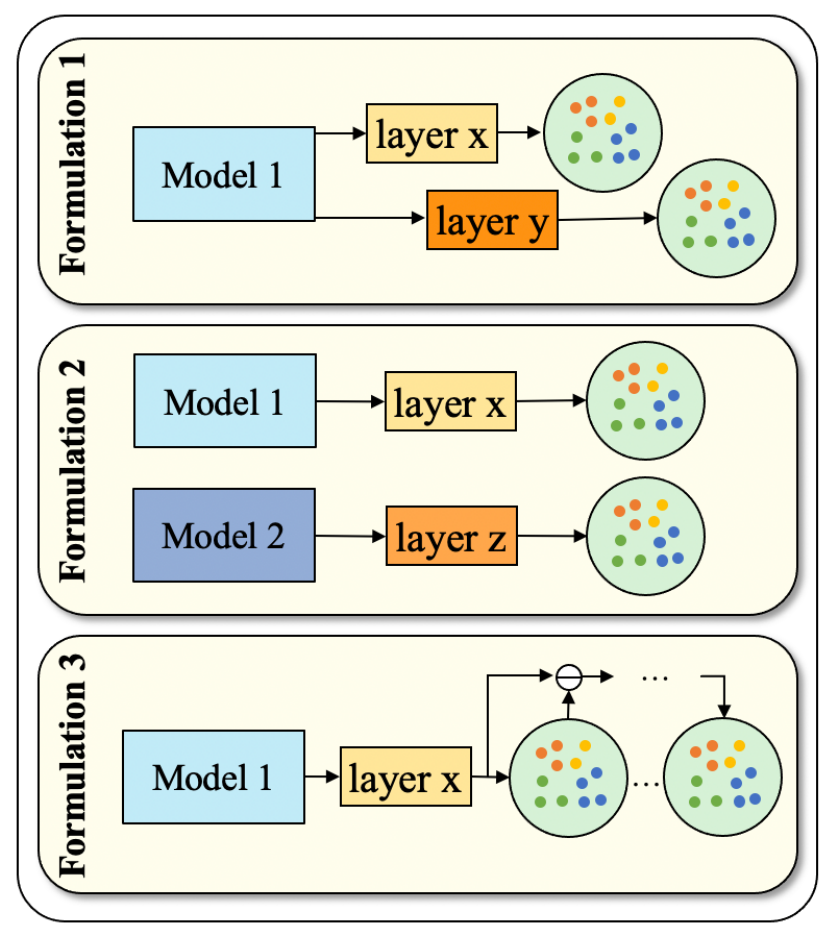}
        \vspace{-8pt}
	\caption{\small Token Formulation 1/2/3 refer to forming from different layers of the same model, blending from different models, and generating from residual quantization, respectively.}
	\label{fig:multi}
\end{figure}

These tokens can serve as intermediate representations, however, the information conveyed by a single token is limited. Inspired by \cite{shi2024Ommm}, we propose a more flexible token formulation that involves blending tokens, which can be categorized into three basic types as in Figure~\ref{fig:multi}: \textbf{\textit{Type 1}} involves selecting tokens from different layers of the same SSL model. \textbf{\textit{Type~2}} contains tokens from different SSL models that may relate to different pre-training corpora and tasks, taking advantages from different SSL models without additional training overhead. Lastly, by using RVQ, multiple-layer tokens can be obtained from the hidden embeddings of SSL models in a residual manner. Alternatively, we can utilize codec tokens directly obtained from audio encoders pre-trained on large-scale corpora~\cite{Zeghidour2021SoundStreamAE, Defossez2022HighFN}. These three basic types can be utilized individually or combined strategically to offer greater flexibility and interpretability.

\subsection{Musical Encoder}
\label{ssec: musical encoder}
The music encoder conducts acoustic modeling of the transition from musical scores to intermediate representations. It takes lyrics and corresponding note sequences containing pitch and duration information as input and outputs intermediate representations at the frame level. SVS requires adherence to the timing variations specified by the given musical scores, which demands higher accuracy in duration prediction. Hence, we employ a non-autoregressive (NAR) model with an explicit duration prediction module for modeling following \cite{ren2019fastspeech, wang2022xiaoicesing, wu2023phoneix}. 

\begin{table*}[!t]
\caption{\small Comparison of the proposed discrete-based system TokSing with a Mel-spectrogram based system \cite{ren2019fastspeech, Lu2020XiaoiceSingAH}, and a latent variance based system \cite{Zhang2022VISinger2H}. The generated audios are evaluated in terms of their performance on objective metrics ({MCD}, {F0}, {SA}) and subjective metrics ({Pron}, {Melody}, {Tech}, {MOS}) described in Section~\ref{ssec: settings}.} 
\vspace{-8pt}
\centering
\resizebox{0.7\textwidth}{!}{
\begin{tabular}{c|ccc|cccc|c}
\toprule
\multirow{2}{*}{\textbf{Representation}} & \multicolumn{3}{c|}{\textbf{Objective Evaluations}} & \multicolumn{4}{c|}{\textbf{Subjective Evaluations}} & \multirow{2}{*}{\textbf{Bitrate/bps}}\\
 & \textbf{MCD ↓} & \textbf{F0 ↓}  & \textbf{SA ↑} & \textbf{Pron ↑} & \textbf{Melody ↑} & \textbf{Tech ↑} & \textbf{MOS ↑} &  \\
\midrule
Mel Spectrogram & 8.00 & 0.19 & 59\% & 2.59  & 2.23 & 2.20 & 3.42  & 204800 \\
Latent Variance  & 7.76  & 0.18 & \textbf{62\%} & \textbf{2.74} & 2.34 & \textbf{2.37} & 3.68 & 491520 \\
Discrete Token & \textbf{7.56} & \textbf{0.17} & 61\% & 2.73 & \textbf{2.38} & \textbf{2.37}  & \textbf{3.70}  & \textbf{1950}\\
\midrule
GT & - & - & - & 2.93 & 2.83 & 2.82 & 4.59  & - \\
\bottomrule
\end{tabular}
}
\label{tab: comparison}
\end{table*}

In addition to the precise timing constraints, singing places a premium on pitch accuracy. As mentioned earlier, the discretization process may entail the degradation of pitch variations in singing, which is validated by our experiments in Section ~\ref{ssec: resynthesis}. To compensate for the loss of melody changes, we include a melody signal in addition to tokens by using the logarithm of the fundamental frequency. To further enhance the accuracy of melody prediction, we introduce a melody encoder to encode the input pitch and utilize a melody predictor for pitch prediction (see purple blocks in Figure ~\ref{fig:system}). 

We compute melody loss function $\mathcal{L}_{m}$ using Euclidean distance between the extracted $m_i$ from the raw audio and the predicted $\widehat{m}_i$ as: 
$\mathcal{L}_{m} = \frac{1}{N} \sum_{i=1}^{N} |m_i - \widehat{m}_i|$, where $N$ represents the number of frames.

The tokens generated by the musical encoder 
might still encapsulate certain pitch-related details. Therefore, the output of the score decoder is enhanced with the predicted $\widehat{m}_i$, jointly passed to the token predictor for better prediction. We compute the token prediction loss $\mathcal{L}_{\text{tok}}$ using the cross-entropy loss as:
$\mathcal{L}_{\text{tok}} = - \frac{1}{N} \sum_{i=1}^{N} \sum_{j=1}^{M} (d_i^j) \log(\hat{d}_i^j)$,
where $d_{ij}$ represents the ground truth token, $\hat{d}_i^j$ represents the predicted token and $M$ is the number of tokens layers.

\begin{figure}[t]
	\centering
	\includegraphics[width=0.8\columnwidth]{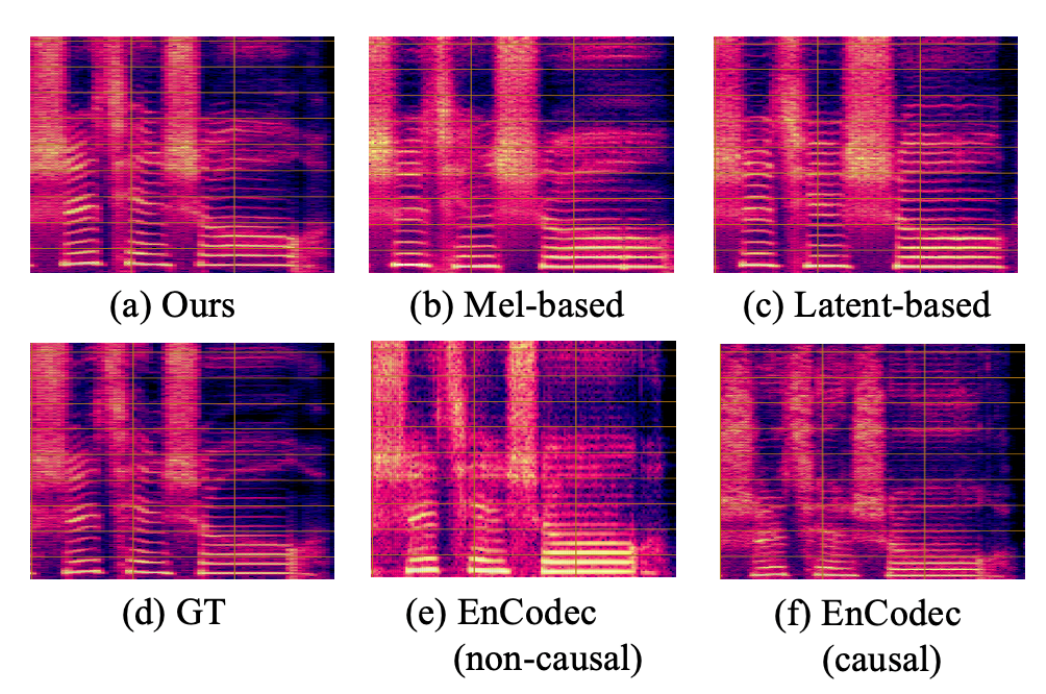}
 \vspace{-8pt}
	\caption{\small Visualization of generated audio segments by different systems. (e) and (f) are resynthesized by codec decoder.}
	\label{fig:case}
 \vspace{-15pt}
\end{figure}

\subsection{Vocoder}
\label{ssec: vocoder}
The backbone of the vocoder adopts the adversarial network from HiFiGAN \cite{kong2020hifi}, comprising a generator and two discriminators: the Multi-Period Discriminator (MPD) and the Multi-Scale Discriminator (MSD). Following previous works \cite{lee2022direct, yan2023espnet}, we replace Mel spectrograms with discrete tokens. The tokens are encoded by extra embedding layers and then concatenated with melody signal, passed to the upsampling layers. Different choices of vocoders are compared in Section ~\ref{ssec: resynthesis}.


\section{Experiment}
\label{sec:exp}

\subsection{Experimental Settings}
\label{ssec: settings}
We conduct experiments in single-singer and multi-singer scenarios. All singing audios have score notations of pitch, duration, and lyrics. During training, audios are preprocessed according to the SSL model's sampling rate and hop size.

\noindent\textbf{\textit{Datasets:}}
We carry out experiments on two public datasets: 1)~
Opencpop~\cite{Wang2022OpencpopAH} comprises 5.2 hours of 100 songs featuring a Mandarin female vocalist. We follow the official split in training and testing sets, with provided sentence-level segmentation. 2) ACE-Opencpop \cite{Shi2024SingingVD} is a dataset derived from Opencpop's music scores, containing  multi-singers synthesized using the ACE Studio\footnote{\scriptsize{\url{https://ace-studio.timedomain.cn}}} with detailed manual tuning. The dataset encompasses 30 singers with diverse genders and vocal styles, accumulating approximately 150 hours of total duration.

\noindent\textbf{\textit{Token Formulation:}}
To acquire suitable token sources, we conduct resynthesis experiments across different hidden layers of various SSL models, aggregating them with a weighted sum approach. We select layers with higher weights as token sources for the following experiments. Eventually, the 6th and 23rd layers of the WavLM-large\footnote{\scriptsize{\url{https://huggingface.co/microsoft/wavlm-large}}} model, as well as the 6th layer of the HuBERT-large model\footnote{\scriptsize{\url{https://huggingface.co/facebook/hubert-large-ll60k}}} are chosen, which also exhibit strong performance in speech understanding tasks~\cite{Yang2021SUPERBSP, shi2023multi}. Additionally, setting the optimal number of clustering centers can be influenced by the phoneme inventory of the language and the number of singers involved. We compare the performance across exponential powers of 2 ranging from 32 to 1024. Ultimately, we set the number of clustering centers to 128 for the single-singer dataset and 1024 for the multi-singer dataset.

\noindent\textbf{\textit{Model configurations:}}
The musical encoder adopts a NAR architecture in \cite{ren2019fastspeech} with an explicit duration predictor. The encoding and decoding of score utilize a transformer based structure following \cite{Lu2020XiaoiceSingAH}, employing a 384-dim embedding layer to encode lyrics, pitches, and durations from music scores. The prediction of melody aligns with \cite{Zhang2022VISinger2H} and the groundtruth melody signals are extracted from raw audios using pyworld \footnote{\scriptsize{\url{https://github.com/JeremyCCHsu/Python-Wrapper-for-World-Vocoder}}} and computed by natural logarithm. The predicted ones are passed through a simple fully connected network (pink block in Figure~\ref{fig:system}) to the token predictor. In vocoder, a 768-dim and a 256-dim embedding layer are used to encode the token and melody signal, respectively. The parameters of the generator and discriminator are consistent with \cite{kong2020hifi}. For the multi-singer dataset, a singer embedding layer is integrated for musical encoder but is not used in vocoder.

\begin{figure}[t]
	\centering
	\includegraphics[width=0.8\columnwidth]{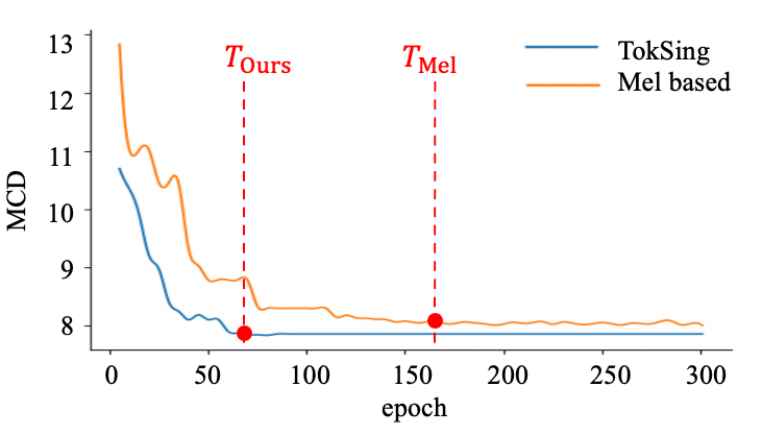}
 \vspace{-8pt}
	\caption{\small Convergence speed comparison of TokSing and the Mel spectrogram-based system.} 
	\label{fig:convergence}
 \vspace{-15pt}
\end{figure}

\noindent\textbf{\textit{Training and Inference:}}
The model employs the Adam optimizer with a learning rate of 0.001. The batch size is set to 16 for training. The inference is performed by averaging the best five models with the lowest loss on the validation set.

\noindent\textbf{\textit{Evaluation Metric:}}
We use common objective metrics, including Mel Cepstral Distortion ({MCD}), Root Mean Square Error of Fundamental Frequency ({F0}), Semitone Accuracy ({SA}). Additionally, we conduct four subjective evaluations, including the clarity in lyric pronunciation ({Pron}), fluency in melody expression ({Melody}), proficiency in singing technique ({Tech}), rated on an integer scale from 1 to 3, followed by an overall listening experience ({MOS}) rating on a scale from 1 to 5. We randomly select 30 identical samples from each system and invite 20 native-speaker annotators to rate them. All annotators undergo pre-annotation tests to ensure they are musically competent. We report the MOS score at a confidence interval of $95\%$.


\subsection{Comparison Experiments}
\label{ssec: comparision}

We compare our discrete-based SVS system, {TokSing}, with a fastspeech-like Mel spectrograms-based system \cite{wang2022xiaoicesing} and a VAE-structured latent variance-based system \cite{Zhang2022VISinger2H}. As mentioned in Section ~\ref{ssec: musical encoder}, singing is more sensitive to the duration variations of notes. Therefore, all systems employ acoustic models with explicit length regulation in a NAR manner. {TokSing} achieves better performance in both subjective and objective metrics, especially in the perception of melody. Cases of the generated segments (see Figure ~\ref{fig:case}) also demonstrate that {TokSing} can exhibit more nuanced performance in melody variations. 

As shown in Figure \ref{fig:convergence}, {TokSing} also shows its advantages in both space cost and convergence speed as it only requires one token and one melody signal to represent a frame, resulting in much lower dimensions than Mel spectrogram-based and latent variance-based systems. Additionally, {TokSing} converges faster and more stably than the system based on Mel spectrograms. We calculate the {MCD} scores on the test set every five epochs. {TokSing} converges ($T_{\text{Ours}}$) significantly earlier than the Mel spectrogram-based system ($T_{\text{Mel}}$).

\subsection{Ablation of the Reconstruction in Vocoder}
\label{ssec: resynthesis}
\begin{table}[!t]
\caption{Ablation of the resynthesis performance of vocoder with different input representation.} 
\vspace{-8pt}
\centering
\resizebox{0.8\columnwidth}{!}{
\begin{tabular}{c|cc|cc}
\toprule
\multirow{2}{*}{\textbf{Representation}} & \multicolumn{2}{c|}{\textbf{Vocoder}} & \multicolumn{2}{c}{\textbf{+ Acoustic}} \\
 & \textbf{MCD ↓} & \textbf{F0 ↓}  & \textbf{MCD ↓} & \textbf{F0 ↓} \\
\midrule
SSL Feat.     & \textbf{3.18}   & \textbf{0.14} & - & - \\
Token-only    & 9.10  & 0.27 & 9.60 & 0.26 \\
\textbf{\textit{Token + $LF0$}} & 6.59 & 0.17  & \textbf{7.56} & \textbf{0.17} \\
\midrule
Codec Token + $LF0$  & 7.56 & 0.17 & - & - \\
Codec Decoder & 6.35 & 0.24 & - & - \\
\midrule
Mel Spectrogram   & 3.32    & 0.15   & 8.00 & 0.19\\
\bottomrule
\end{tabular}
}
\label{tab: vocoder}
\end{table}
\vspace{-4pt}

We note in Section~\ref{ssec: musical encoder} that discretization may entail melody degradation. We conduct reconstruction experiments from the hidden layer of SSL models and from tokens extracted using the same layer. Results in Table \ref{tab: vocoder} indicate a significant decrease in melody quality when using only tokens. 
To mitigate the melody loss during discretization, we introduce a melody signal described in Section~\ref{ssec: musical encoder} as an additional auxiliary representation. The results validate the effectiveness of this strategy both in the vocoder and when connected to the musical encoder. 
Although discrete tokens may not outperform Mel spectrogram-based vocoders in reconstruction experiments, they contribute to better overall performance in the SVS system. This suggests that discrete representations alleviate the modeling difficulty in the acoustic model. Compared to vocoders trained directly on the intermediate layers of self-supervised models, discrete-based vocoders still have room for further improvement.

Additionally, we test the reconstruction quality by training a new vocoder using codec tokens and those reconstructed by the codec decoder. For the former, we select the first layer of codec tokens from Encodec (non-causal)~\cite{Defossez2022HighFN} (referred to as Codec Token + $LF0$ in Table\ref{tab: vocoder}). For the latter, we follow the encoder-decoder codec architecture of the same model to reconstruct singing audios (referred to as Codec Decoder in Table~\ref{tab: vocoder}). Examples of more codec decoder reconstructions can be found in Figure~\ref{fig:case} (e) and (f). The experiments indicate that the performance of codec tokens is inferior to tokens obtained from the SSL model. Furthermore, the codec decoder performs poorly in pitch accuracy.


\subsection{Ablation of Musical Encoder.}
\label{ssec: melody}
\begin{table}[!t]
\caption{Ablation of the melody predictor and melody enhancement described in Section~\ref{ssec: musical encoder} in musical encoder.} 
\vspace{-8pt}
\centering
\resizebox{1\columnwidth}{!}{
\begin{tabular}{cc|ccc|cc}
\toprule
\textbf{Melody} & \textbf{Melody} & \multicolumn{3}{c|}{\textbf{Objective}}  & \multicolumn{2}{c}{\textbf{Subjective}} \\
\textbf{Prediction} & \textbf{Enhanced} & \textbf{MCD ↓} & \textbf{F0 ↓} & \textbf{SA ↑} & \textbf{Melody ↑} & \textbf{MOS ↑} \\
\midrule
\XSolidBrush  & \XSolidBrush & 8.07  & 0.19 & 44\% & 2.07 & 3.18 \\
\XSolidBrush &\Checkmark & 7.61  & 0.21  & 59\% & 2.32 & 3.65 \\
\Checkmark &\Checkmark & \textbf{7.56}  & \textbf{0.17} & \textbf{61\%} & \textbf{2.38} & \textbf{3.70} \\
\bottomrule
\end{tabular}
}
\label{tab: melody}
\end{table}
\vspace{-2pt}
Section~\ref{ssec: musical encoder} mentions that 
tokens may also carry some melody information. Therefore, in addition to introducing a melody predictor, we employ a melody enhancement strategy to assist in token prediction (implementation details in Section~\ref{ssec: settings}). The ablation experiments presented in Table~\ref{tab: melody} demonstrate that both the melody predictor and the melody enhancement strategy enhance the performance of SVS system, particularly in terms of melody expression.

\subsection{Ablation of Token Formulations}
\label{ssec: type}

\begin{table}[!t]
\caption{Ablation of the SVS performance under different token formulations described in Section~\ref{ssec: token formulation}. }
\vspace{-8pt}
\centering
\resizebox{0.85\columnwidth}{!}{
\begin{tabular}{c|cc|ccc}
\toprule
\multirow{2}{*}{\textbf{Representaion}} & \multicolumn{2}{c|}{\textbf{Vocoder}} & \multicolumn{3}{c}{\textbf{+ Acoustic}} \\
 & \textbf{MCD ↓} & \textbf{F0 ↓}  & \textbf{MCD ↓} & \textbf{F0 ↓} & \textbf{MOS ↑} \\
\midrule
Single    & 6.59  & 0.17 & 7.56 & \textbf{0.17} & 3.70 \\
\midrule
\textbf{\textit{Formulation 1}}     & 6.51  & 0.17 & 7.59 & 0.18 & \textbf{3.74} \\
\textbf{\textit{Formulation 2}}     & 6.39  & 0.16 & \textbf{7.50} & \textbf{0.17} & 3.61\\
\textbf{\textit{Formulation 3}}  & \textbf{5.96}  & \textbf{0.15} & 7.65 & 0.18 & 3.40 \\
\bottomrule
\end{tabular}
}
\label{tab: token}
\end{table}

We compare the performance of single-layer tokens with the three formulations of multi-layer token formulations mentioned in Section~\ref{ssec: token formulation}. The single-layer tokens originate from the 6th layer of WavLM-large. 
For multi-layer token formulations, {Formulation 1} incorporates tokens from the 23rd layer of the same model; {Formulation 2} adds the 6th layer from a different model, HuBERT-large; and {Formulation 3} applies a four-layer residual clustering. Both subjective and objective evaluations indicate that all multi-layer formulations enhance performance to varying degrees. It is noteworthy that although residual clustering in {Formulation 3} improves the performance of the vocoder by encoding more details, it also increases the prediction difficulty in the musical encoder. We observe a noticeable decrease in prediction accuracy in the later layers of residual clustering.








\subsection{Transfer Learning}
\label{ssec: transfer}

Experiments in Section~\ref{ssec: resynthesis} suggest potential for improving the token-based vocoder, compared to the vocoder trained on SSL model hidden layers. In Table~\ref{tab: transfer}, we explore transfer learning to enhance vocoder, which is first pretrained on ACE-Opencpop, and then finetuned on Opencpop. The results indicate that transfer learning can further improve  vocoder.

\begin{table}[!t]
\caption{\small Transfer learning from multi singer dataset ACE-Opencpop to single singer dataset Opencpop. }
\vspace{-8pt}
\centering
\resizebox{0.8\columnwidth}{!}{
\begin{tabular}{c|cc|cc}
\toprule
\multirow{2}{*}{\textbf{Representation}} & \multicolumn{2}{c|}{\textbf{Vocoder}} & \multicolumn{2}{c}{\textbf{+ Acoustic}} \\
 & \textbf{MCD ↓} & \textbf{F0 ↓}  & \textbf{MCD ↓} & \textbf{F0 ↓}  \\
\midrule
ACE-Opencpop    & \textbf{5.60}  & \textbf{0.14} & - & - \\
Opencpop-origin & 6.51 & 0.16 & 8.15 & 0.19 \\
Opencpop-transfer & 6.11 & 0.16 & \textbf{7.43} & \textbf{0.17} \\
\bottomrule
\end{tabular}
}
\label{tab: transfer}
\end{table}
\vspace{-8pt}

\section{Conclusion}

This paper proposes a SVS framework TokSing based on discrete representations. 
A melody enhancement strategy is introduced to compensate for the loss of melody information during the discretization process. Experiments demonstrate that the discrete tokens are more efficient for singing representation and achieves superior singing synthesis quality with lower spatial overhead compared to traditional continuous representations.

\section{Acknowledgements}
This work was partially supported by the  National Natural Science Foundation of China (No. 62072462) and the Beijing Natural Science Foundation (No. L233008).


\section{References}
{
\printbibliography
}

\end{document}